\documentclass[prl,twocolumn,amsmath,amssymb,10pt,superscriptaddress,floatfix,showpacs]{revtex4}
\usepackage{latexsym,color,epsfig}

\def\eff{_{\rm eff}}

\def\eref#1{Eq.\thinspace\ref{#1}}
\def\erefs#1{Eqs.\thinspace\ref{#1}} 

\def\fref#1{Fig.~\ref{#1}}
\def\Fref#1{Figure~\ref{#1}}

\def\kT{k_{\rm B}T}
\newcommand{\NR}{N_{\rm R}}
\newcommand{\erf}{\text{erf}}

\def\pnlabel#1{\label{#1}}

\begin{document}

\title{Excluded-Volume Effects in Tethered-Particle Experiments: Bead
Size Matters}

\author{Darren E. Segall}
\affiliation{Division of Engineering and Applied Science,
California Institute of Technology, Pasadena, CA 91125 USA}

\author{Philip C. Nelson}
\affiliation{Department of
Physics and Astronomy, University of Pennsylvania, Philadelphia PA
19104, USA}

\author{Rob Phillips}
\affiliation{Division of Engineering and Applied
Science and Kavli Nanoscience Institute, California Institute of
Technology, Pasadena, CA 91125 USA}

\begin{abstract}
The tethered-particle method is a single-molecule technique that has
been used to explore the dynamics of a variety of macromolecules of
biological interest.  We give a theoretical analysis of the particle
motions in such experiments. Our analysis reveals that the proximity
of the tethered bead to a nearby surface (the microscope slide) gives
rise to a volume-exclusion effect, resulting in an entropic force on
the molecule. This force stretches the molecule, changing its
statistical properties.  In particular, the proximity of bead and surface brings about intriguing scaling relations between key
observables (statistical moments of the bead) and parameters such
as the bead size and contour length of the molecule.  We present
both approximate analytic solutions and numerical results for these effects in both flexible
and semiflexible tethers.  Finally,
our results give a precise, experimentally-testable prediction for the
probability distribution of the distance between the polymer
attachment point and the center of the mobile bead. 
\end{abstract}
\maketitle

\noindent Single-molecule biophysics has rapidly become an
experimental centerpiece in the dissection of  cellular machinery.
This part of the biophysics repertoire often relies, in
turn, on the use of micron-scale beads both as a reporter of
underlying molecular motions and as the ``handle'' for grabbing
these single-molecule systems.  Thus, a key part of the
theoretical infrastructure of this field is a clear understanding
of the role that these beads play in altering the statistical
properties of the macromolecules which are the real target of
interest in such experiments.

Beyond interest in the {\it in-vitro} consequences of tethered-particle motions, many processes within the cell themselves involve
tethering.  A notable example has to do with vesicular trafficking, in
which molecular motors~\cite{hirokawa05} carry tethered cargoes with
similar Earth-like proportions relative to the molecular Atlases doing
the heavy lifting.  Thus, the statistical-mechanical analysis
performed here may prove useful for understanding {\it in-vivo}
processes, in addition to the {\it in-vitro} consequences that form
the main motivation for the work.

\Fref{TPMSchematic} sketches the tethered particle method (TPM).  The
main idea is that a macromolecule (for example DNA or some protein
that translocates DNA or RNA) is anchored at one end to a surface,
while the other end of the molecular complex is attached to a
bead. The observed motion of the bead serves as a reporter of the
underlying macromolecular motion. This technique has been used in a
variety of settings {\it e.g.}\ the examination of nanometer-scale
motions of motors like kinesin~\cite{gelles88} or RNA
polymerase~\cite{schafer91,yin94}, protein synthesis by
ribosomes~\cite{vanzi03}, exonuclease translocation on
DNA~\cite{dohoney01,oijen03}; protein mediated deformation
\cite{dixi05a} and loop formation~\cite{finzi95} in DNA, DNA
hybridization~\cite{singh_zocchi03} and DNA
motion~\cite{pouget04,brog05a}. The main goal of this paper is to show
how the proximity of the reporter bead to the surface affects the
interpretation of the reported data and can even alter the
conformation of the macromolecule of interest.  A theoretical
understanding of these effects will improve the ability to use the TPM
for quantitative~\cite{fn1} analysis of biomolecular properties at the
single molecule level.

\begin{figure}
\begin{center}
\centerline{\epsfxsize=3.4truein \epsfbox{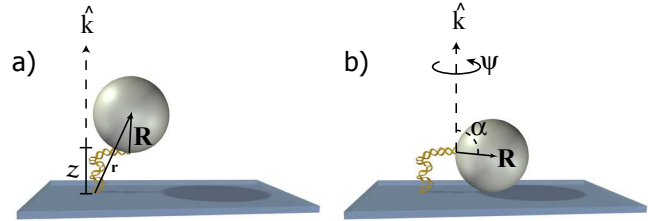}}
\caption[]{\pnlabel{TPMSchematic}Schematic showing the tethered
particle method. {(a)} The tether is attached to a specific point on
the bead; $z$ is the height of this point. $\mathbf{r}$ denotes the
position of the center of the bead. {(b)} The vector $\mathbf{R}$ from
the attachment point to the bead center can rotate and is described by
two angles.  These rotations are more constrained for small values of
$z$.  Note that in the figure the width of DNA is not to scale, it is much
smaller in real experiments.}
\end{center}
\end{figure}

In the remainder of the paper, we first describe a simple statistical-mechanical
theory of bead-induced volume-exclusion forces.  We show how these
forces depend both upon bead size and on tether length.  We also derive
 scaling relations between the experimental measurables (bead
position) and parameters such as bead size and tether length.  Because the simple analytic model neglects some
features of the full problem, we then turn to simulation results which
capture all of the key effects and compare to the analytic results.

The aim of the calculations outlined below is to illustrate how the
presence of a bead alters the statistical properties of the molecule
to which it is tethered and how the bead reports information to the
experimenter.  We confine our discussion to the equilibrium
characteristics of this phenomenon, a key measurable in TPM
experiments even for the study of dynamical
processes~\cite{schafer91,yin94,vanzi03,dohoney01,oijen03,finzi95,pouget04}.
We first note that in many experiments, the bead is flexibly linked to
the end of its molecular tether, and hence is nearly free to rotate
around the point of linkage \cite{yin94,brog05a}. However, steric
constraints limit this freedom.  In particular, the closer the bead is
to the surface, the fewer angular conformations are available to it
(\fref{TPMSchematic}).

The statistical properties of the bead in TPM are determined by the
coarse-grained free energy function (or ``Hamiltonian'')
\begin{equation}
H = H_m(\{X\}) + H_{b,m}(\mathbf R,\{X\})  +
H_b(\mathbf R,\{X\}).  \pnlabel{eq:H}
\end{equation}
Here $\{X\}$ is an abstract set of coordinates describing the
configuration of the molecule and $\mathbf R$ is the vector pointing
from the end-point of the molecule to the center of the bead
(\fref{TPMSchematic}). \eref{eq:H} contains three terms: The first
describes the self-interactions of the molecule and interactions with
external forces (surface forces or applied fields) other than those
with the bead itself. Those interactions are captured by the second
term, $H_{b,m}$.  The last term, $H_b$, describes the external forces
on the bead, for example those arising from applied fields or the
surface. This term also depends on the configuration of the molecule,
as the position of the bead depends on both its orientation $\mathbf
R$ and the molecule's end-point.

We obtain the statistical average of an observable $A$
of the system as a weighted average over all the
configurations of  the system
\begin{equation*}
\langle A\rangle= \frac{1}{Z}\int d\{X\}
d^2\hat{\mathbf R} \
A(\{X\},\mathbf R)\ e^{-\beta H}.
\pnlabel{eq:av}
\end{equation*}
Here $\int d^2\hat{\mathbf R}$ is an integral over bead orientations
and $Z$ is the partition function.

Before discussing the consequences of this model, we first discuss the
significance of the terms in \eref{eq:H} and make some initial
simplifications. The external forces acting on the bead that we have
in mind result from its interaction with the surface. This interaction
contains the repulsive double layer potential and an attractive van
der Waals interaction~\cite{israelachvili91}, along with a hard-wall
repulsion (\eref{eq:Hbmb} below). Under physiological conditions, the
double layer potential has an interaction range with a typical length
scale of a nanometer, much shorter than the molecule lengths of
interest to us; the van der Waals attraction, too, is weak on long
scales~\cite{dagastine04}.  Accordingly, we will model all bead--wall
interactions using  a simple hard-wall potential. We will also temporarily ignore
bead--molecule interactions; later we will show that they have little
influence for tethers like the ones of interest here.  Given these
assumptions, the last two terms in \eref{eq:H} simplify to
\begin{equation}
H_{b,m} = 0, \hspace{1cm} H_b = \left\{ \begin{array}{cl} 0 & \text{if
} R(1-\cos(\alpha)) < z\\ \infty & \text{if } R(1-\cos(\alpha)) \geq
z,
\end{array} \right. \pnlabel{eq:Hbmb}
\end{equation}
where $z$ is the height of the end-point of the molecule and $\alpha$ is the
polar angle of $\mathbf R$ (\fref{TPMSchematic}). 

We now examine the statistical averages of a molecular property
$A_m(\{X\})$ such as end-to-end distance.  To obtain the average value
of $A_m(\{X\})$, not only do we need to sum over all of the
configurations of the molecule, but also, we must sum over all of the
configurations of the bead. Thanks to the simplifications in
\eref{eq:Hbmb}, the integration over $\mathbf R$ can be done
explicitly and results in an {\it effective} free energy function for
the molecule.  That is, the resulting statistical average of $A_m$ can
be written as
\begin{eqnarray}
\langle A_m\rangle & = & \frac{1}{Z}\int d\{X\} A_m(\{X\})e^{-\beta
H\eff},\mbox{\ where} \pnlabel{eq:aveeff}\\  
H\eff & = & H_m(\{X\}) - \kT\log(\Omega(z)).
\pnlabel{eq:Heff}
\end{eqnarray}
The new second term in $H\eff$ accounts for the configurations available to the
bead. $\Omega(z)$  is
the solid angle allowed for $\mathbf R$, given a molecular
configuration $\{X\}$:
\begin{equation}
\Omega(z) = \left\{ \begin{array}{cl} 2\pi{z}/{R}, & z < 2R \\
4\pi, & z \geq 2R. \end{array}\right. \pnlabel{eq:omega}
\end{equation}
The partition function is also consistent with the definition
of the effective Hamiltonian, $Z = \int d\{X\} e^{-\beta
H\eff}$.

As a result of the constraints on the excursions of the bead there is
an effective repulsive force, which prevents the end-point of the
molecule from making contact with the surface and stretches the
molecule.  That is, the problem is equivalent to one without the bead,
but in which the end-point of the molecule is subjected to a
force
\begin{equation}
\mathbf{F}\eff ={z}^{-1} \Theta(2R-z){\kT}\ \hat{\mathbf{k}},
\pnlabel{eq:Feff}
\end{equation}
where $\Theta$ is the Heaviside step function.  This entropic force
alters the statistical properties of the tethered molecule
(\eref{eq:aveeff}), and can affect its interactions with itself or
with other molecules.

The key measurable associated with current TPM experiments is the
position of the bead itself. That is, the output of the experiment is
a record of the positions of the bead on successive video
frames~\cite{vanzi03,pouget04,brog05a,tpm05}.  Having revealed that the
confinement of the bead subjects the molecule to an entropic force, we
now determine how this confinement influences bead motion.  Note that one of our key conclusions
is that there is a subtle dependence of the measured bead excursions
on the size of the bead itself.  To see this, let $\mathbf r$ (\fref{TPMSchematic}) denote the vector
from the wall attachment point to the bead center.  Given our
simplifications (\eref{eq:Hbmb}), the non-zero statistical moments of
$\mathbf r$, up to second order, are
\begin{subequations}
\pnlabel{eq:rstats}
\begin{eqnarray}
\!\!\!\!\!\!\!\!\langle r_z\rangle &\!=& \!\langle z\rangle -
{\textstyle\frac{1}{2}}\langle z\rangle_{\Theta} + R\langle
1\rangle_{\Theta} \label{eq:rstatsrz} \\ \!\!\!\!\!\!\!\!\langle
{r_z}^2 \rangle &\!=& \!\langle z^2 \rangle - {\textstyle\frac{2}{3}}
\langle z^2\rangle_{\Theta} + R\langle z\rangle_{\Theta} +
R^2({\textstyle\frac13} +{\textstyle \frac23}\langle
1\rangle_{\Theta}) \label{eq:rstatsrz2}\\ \!\!\!\!\!\!\!\!\langle
{\mathbf r_\perp}^2 \rangle &\!=& \!\langle {\mathbf x_\perp}^2\rangle
- {\textstyle\frac{1}{3}}\langle z^2\rangle_{\Theta} + R\langle
z\rangle_{\Theta} +{\textstyle \frac23} R^2(1-\langle
1\rangle_{\Theta})
\label{eq:rstatsrperp2}
\end{eqnarray}
\end{subequations}
where ${\mathbf r_\perp}$ is the in-plane displacement of the bead
center and $({\mathbf x_\perp},z)$ is the displacement of the
end-point of the molecule.  Here all quantities averaged on the
right-hand side are independent of the coordinates of the bead and,
hence, are defined as in~\eref{eq:aveeff}.  The averages with
subscript $\Theta$ correspond to summing over states only when the
end-point satisfies $z < 2R$, {\it i.e.}
\begin{equation}
\langle A_m\rangle_\Theta = \frac{1}{Z}\int
d\{X\}A_m(\{X\})e^{-\beta \rm H_{\rm eff}}\Theta(2R-z).
\end{equation}
These $\Theta$-weighted terms arise because of the reduction of the
configurations available to the bead due to the proximity of the
surface (\eref{eq:omega}).  Below it is shown how they give rise to
experimentally testable scaling relations relating the excursion of
the bead to its radius and the contour length of the molecule.

We further pursue an analytic description of TPM by modeling the
molecule as a Gaussian chain.  The Gaussian chain is a useful
approximation for molecules with short persistence lengths (such
as RNA) and we demonstrate, using numerical simulations, that it
also serves as a good guide for semi-flexible molecules ({\it
e.g.}\ DNA) in the regime of interest here.

Our problem involves a molecule grafted onto a surface, {\it i.e.}\
confined to a half space.  DiMarzio showed that the corresponding
Gaussian chain is described by~\cite{dimarzio65}
\begin{equation}
H_{\rm m} = ({3\kT}/{4L\xi})({\mathbf x_\perp}^2 + z^2) -
\kT\log(z/\ell).\pnlabel{eq:HmG}
\end{equation}
$L$ is the contour length of the molecule, $\xi$ is the
persistence length and $\ell$ is an arbitrary constant.  The material
properties of the molecule enter $H_{\rm m}$ only through the
combination $L\xi$.

Because there are only two relevant length
scales in the problem ($R$ and $\sqrt{L\xi}$),
we may write each moment of the molecule's excursions in terms of a 
function of a single
variable. Using \eref{eq:HmG} to evaluate the averages in
\erefs{eq:rstats}:
\begin{subequations}
\label{eq:rgaus}
\begin{eqnarray}
\frac{\langle r_z\rangle}{\sqrt{L\xi/3}} &=&
\frac{2(1 - e^{-\NR^2})}{\sqrt{\pi}\erf(\NR)} + \NR\frac{2 -
\erf(\NR)}{\erf(\NR)}, \\ \frac{\langle {r_z}^2\rangle}{L\xi/3} &=&
2 + \frac{4\NR}{\sqrt{\pi}\erf(\NR)} + \NR^2,\\ \frac{\langle
{\mathbf r_\perp}^2\rangle}{L\xi/3}  &=& 2 +
\frac{4\NR}{\sqrt{\pi}\erf(\NR)}.
\end{eqnarray}
\end{subequations}
The excursions depend on the dimensionless number $\NR\equiv
{R}/{\sqrt{L\xi/3}}$, which we call the ``excursion number.''  $\NR$
controls the bead's scaling behavior, defining regimes of
molecule-dominated motion ($\NR < 1$) and bead-dominated motion ($\NR
> 1$, confined rotations).
\begin{figure}
\begin{center}
\centerline{\epsfxsize=2.4truein \epsfbox{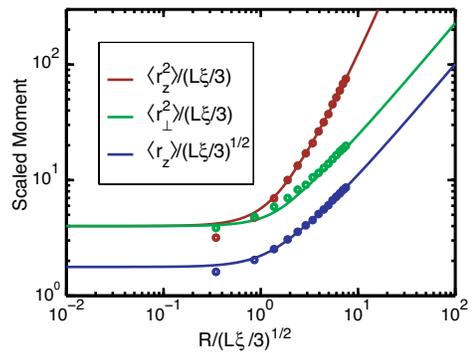}}
\caption{\pnlabel{fig:scaling} Scaling behavior of bead excursion,
normalized by coil size parameter, versus the excursion number $\NR$.
{\it Curves:} analytical theory in the Gaussian-chain approximation
(\eref{eq:rgaus}). {\it Circles:} Monte Carlo calculation for a
semiflexible chain with $\xi=50\,$nm, $L=1245\,$bp, and various values
of $R$.  }
\end{center}
\end{figure}

\Fref{fig:scaling} shows the relationship between excursions and
$\NR$.  For small excursion number, the excursions scale as $\langle
{r_z}^2\rangle \approx \langle {\mathbf r_\perp}^2\rangle \approx
L\xi$ and $\langle r_z\rangle \approx \sqrt{L\xi}$; the dependence on
contour length obeys the expected relations for a Gaussian chain with
no bead attached.  The scaling changes for large excursion number.
Now, the mean excursions display power-law dependences on $\NR$:
$\langle {\mathbf r_\perp}^2\rangle \approx R\sqrt{L\xi}, \langle
{r_z}^2\rangle \approx R^2$ and $\langle r_z\rangle \approx R$; the
observed motion is dominated by the bead's rotation.  The power law
difference on bead radius $R$ follows directly from the general
relations~(\eref{eq:rstats}), independent of the model representing
the molecule.  Now, the in-plane excursions show a square-root
dependence on the contour length, in contrast with the linear
dependence for small excursion number. This functional form arises
because the average height of the molecule (which depends on
$\sqrt{L\xi}$) dictates the degree to which the bead can rotate.
These scaling relations should be testable in experimental studies,
where excursion numbers have ranged over wide set of values, $0.1\leq
\NR \leq
90$~\cite{schafer91,yin94,vanzi03,dixi05a,finzi95,singh_zocchi03,pouget04}.
(Most are in the regime $\NR > 1$.)
\begin{figure}[t!]
\begin{center}
\centerline{\epsfxsize=2.4truein \epsfbox{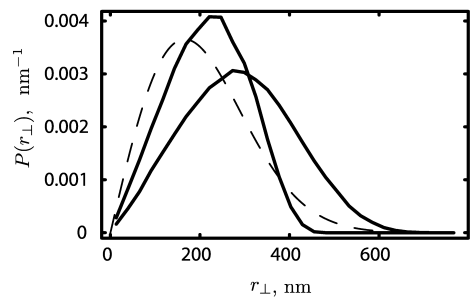}}
\caption[]{{\it
Solid curves:} Theoretical prediction of the probability
distributions for the projected distance
${\mathbf r_\perp}$, taking bead radius $R=250\,$nm, persistence length
$\xi=50\,$nm, and contour length $L=1000\,$bp (left curve) and 2000
(right curve). {\it Dashed curve:} Two dimensional Gaussian
distribution with the same mean-square excursion as the left curve.
\pnlabel{f:phil}}
\end{center}
\end{figure}

Confinement effects can alter molecular properties, causing
out-of-plane stretching of the
molecule. \erefs{eq:aveeff}~and~\ref{eq:HmG} yield:
\begin{equation}
   \frac{\langle z^2\rangle}{L\xi/3} = 6 -
\frac{4}{\sqrt{\pi}}\frac{\NR e^{-\NR^2}}{\erf(\NR)}.\pnlabel{eq:z2}
\end{equation}
In contrast, a Gaussian chain in free solution has squared
out-of-plane excursions of the molecule equal to $\langle
z^2\rangle/(L\xi/3) = 2$.  For many
experiments~\cite{yin94,vanzi03,dohoney01,dixi05a,finzi95,singh_zocchi03,pouget04}
the excursion number is such that the second term in~\eref{eq:z2} is
negligible and the effects of bead exclusion (\eref{eq:Feff}) and
molecule exclusion (second term in~\eref{eq:HmG}) result in a tripling
of the out-of-plane squared displacement of the molecule.  (Both
exclusion effects contribute equally.)  The bead-induced stretching 
can be viewed
as a consequence of the effective force
(\eref{eq:Feff}).  In the Gaussian chain model, this force is
\begin{equation}
\langle F\eff\rangle = \frac{\kT}{\sqrt{\pi}\sqrt{L\xi/3}}\left(\frac{1
- e^{-\NR^2}}{\erf(\NR)}\right).\pnlabel{eq:Feffgaus}
\end{equation}
This force can significantly affect rates of loop formation in DNA.
Finzi and Gelles~\cite{finzi95} used TPM to observe loop formation in
DNA generated by the lac-repressor protein.  Under their conditions,
we predict an average effective force $\langle F\eff\rangle \approx
25$~fN.  Using the simple approximation that the rate of loop
formation decreases by $\exp(-\beta \langle F\eff\rangle l)$ ($l$ is
the operator-operator distance) we estimate that the bead-confinement
effect reduces the rate of loop formation by a factor of 2.

To check the validity of our simplified model, we performed a simple
Monte Carlo calculation. Our results agree with an independent
calculation by D.~Brogioli \cite{brog05a}. Our code generated sets of
discrete chains with random bends chosen to obtain a desired
persistence length $\xi$. Each chain began at a random angle relative
to the wall, and ended with the bead at a random
orientation. Configurations where the bead, wall, or chain overlapped
were discarded, (this includes bead-molecule interactions ($H_{b,m}\neq
0$)) and the required averages were computed.  \Fref{fig:scaling}
shows that even for a stiff polymer like DNA, the scaling relations
predicted by the approximate analytical theory are accurate in the
regime of interest to us.  Actual experimental data allow the
calculation of more subtle metrics than just averages, however:
\fref{f:phil} shows predictions for the full probability distribution
of excursions.  The distribution is quite different from a Gaussian, a
fact already observed experimentally~\cite{pouget04}.

The above results are interesting as
fundamental polymer physics. For example, single-particle tracking
allows the observation of the full probability distribution, and hence
the opportunity to directly observe an end-end distribution for a
semiflexible polymer and compare to our predictions. But our main goal
was to develop a theoretical framework which can bolster the
quantitative capabilities of the TPM---a relatively noninvasive,
single molecule probe.  We revealed that the proximity of the bead to
the surface provokes an effective force on the molecule, altering its
statistical properties and influencing biomolecular interactions.  In
addition, we determined how the excursions of the bead are influenced
by experimental parameters such as bead size and contour lengths; relations
which are currently being tested~\cite{tpm05}.  Finally, understanding
how the competition between bead and tether effects is controlled by
the excursion number $\NR$ may help in the choice of 
optimal bead size and tether length for a given experiment.

\begin{acknowledgments}
We thank J. Beausang, S. Blumberg, D. Brogioli, D. Chow, D. Dunlap,
L. Finzi, J. Gelles, Y. Goldman, I. Kulic, J.C. Meiners, T. Perkins,
P. Purohit, F. Vanzi, P. Wiggins, C. Zurla, for extensive discussions.
DS acknowledges the support of AFOSR/DARPA grant F49620-02-1-0085.  RP
acknowledges the support of the NIH Director's Pioneer Award DP1
OD000217 and NSF grant CMS--0301657. PN acknowledges the Human
Frontier Science Foundation and NSF grant DMR--0404674.
\end{acknowledgments}

\bibliography{tpm}

\begin{thebibliography}{10}

\bibitem{hirokawa05}
N.~Hirokawa and R.~Takemura.
\newblock {\em Nat. Rev. Neurosci.}, 6:201, 2005.

\bibitem{gelles88}
J.~Gelles, B.~J. Schnapp, and M.~P. Sheetz.
\newblock {\em Nature}, 331:450, 1988.

\bibitem{schafer91}
D.~A. Schafer, J.~Gelles, M.~P. Sheetz, and R.~Landick.
\newblock {\em Nature}, 352:444, 1991.

\bibitem{yin94}
H.~Yin, R.~Landick, and J.~Gelles.
\newblock {\em Biophys. J.}, 67:2468, 1994.

\bibitem{vanzi03}
F.~Vanzi, S.~Vladimirov, C.~R. Knudsen, Y.~E. Goldman, and B.~S. Cooperman.
\newblock {\em RNA}, 9:1174, 2003.

\bibitem{dohoney01}
K.~M. Dohoney and J.~Gelles.
\newblock {\em Nature}, 409:370, 2001.

\bibitem{oijen03}
A.~M. van Oijen, P.~C. Blainey, D.~J. Crampton, C.~C. Richardson,
  T.~Ellenberger, and X.~S. Xie.
\newblock {\em Science}, 301:1235, 2003.

\bibitem{dixi05a}
S.~Dixit, M.~Singh-Zocchi, J.~Hanne, and G.~Zocchi.
\newblock {\em Phys. Rev. Lett.}, 94:118101, 2005.

\bibitem{finzi95}
L.~Finzi and J.~Gelles.
\newblock {\em Science}, 267:378, 1995.

\bibitem{singh_zocchi03}
M.~Singh-Zocchi, S.~Dixit, V.~Ivanov, and G.~Zocchi.
\newblock {\em Proc. Natl. Acad. Sci. USA}, 100:7605, 2003.

\bibitem{pouget04}
N.~Pouget, C.~Dennis, C.~Turlan, M.~Grigoriev, M.~Chandler, and L.~Salom\'{e}.
\newblock {\em Nucl. Acids Res.}, 32:e73, 2004.

\bibitem{brog05a}
D. Brogioli, C. Zurla, P. C. Nelson, D. D. Dunlap and L. Finzi, in preparation.

\bibitem{fn1}
Earlier theoretical work {\cite{qian99}} discussed an earlier version of the
  experimental technique, in which only the blurred image of the bead was
  observed. However, in that work bead-wall exclusion effects were neglected,
  critical for the analysis of tethered-particle experiments. In addition,
  recent experiments combine a tethered particle with single-particle tracking
  {\cite{vanzi03,pouget04,brog05a,tpm05}}, requiring analysis such as that
  presented here.

\bibitem{israelachvili91}
J.~Israelachvili.
\newblock {\em Intermolecular and Surface Forces}.
\newblock Academic Press, London, 1991.

\bibitem{dagastine04}
R.~R. Dagastine, M.~Bevan, L.~R. White, and D.~C. Preive.
\newblock {\em J. Adhesion}, 80:365, 2004.

\bibitem{tpm05}
J. Gelles, J.C. Meiners, L. Finzi and R. Phillips (private communications).

\bibitem{dimarzio65}
E.~A. DiMarzio.
\newblock {\em J. Chem. Phys.}, 42:2101, 1965.

\bibitem{qian99}
H.~Qian and E.L. Elson.
\newblock {\em Biophys. J.}, 76:1598, 1999.

\end{thebibliography}

\end{document}